\documentclass{article}
\newenvironment{keywords}{\begin{quote}{\bf Keywords:}}{\end{quote}}
\newcommand\PARstart[2]{#1#2}
\newenvironment{proof}{{\sl Proof:}}{\hfill{$\square$}}
\newcommand\IEEEmembership[1]{\ignorespaces}

\usepackage{amssymb}
\usepackage{amsmath}
\usepackage{pstricks}
\usepackage{graphics}

\newcommand\mut[1]{\ignorespaces}

\newtheorem{theorem}{Theorem}[section]

\newtheorem{lemma}[theorem]{Lemma}
\newtheorem{remarkit}[theorem]{Remark}
\newtheorem{exampleit}[theorem]{Example}

\renewcommand{\leq}{\leqslant}
\renewcommand{\geq}{\geqslant}
\def\N{{\mathbb N}}
\def\Np{{\mathbb N}}
\def\Nz{{\mathbb N}_0}
\def\Nmin{{\mathbb N}_{-1}}
\def\Z{{\mathbb Z}}
\def\P{{\mathbb P}}
\def\F{{\mathbb F}}
\def\Fq{{\mathbb F}_q}
\def\mB{{\mathcal B}}
\def\mF{{\mathcal F}}
\def\mL{{\mathcal L}}
\def\mM{{\mathcal M}}
\def\mN{{\mathcal N}}
\def\mO{{\mathcal O}}
\def\mT{{\mathcal T}}
\def\phi{\varphi}
\def\pless{\prec}
\def\pleq{\preccurlyeq}
\def\pgr{\succ}
\def\pgeq{\succcurlyeq}
\def\pplus{\oplus}
\def\pmin{\ominus}
\def\rta{\rightarrow}
\def\lta{\leftarrow}
\def\lrta{\leftrightarrow}
\def\Rta{\longrightarrow}
\def\Lta{\longleftarrow}
\def\Equiv{\Longleftrightarrow}
\newcommand{\mrta}[1]{\stackrel{#1}{\rightarrow}}
\newcommand{\mRta}[1]{\stackrel{#1}{\longrightarrow}}
\def\Goto{\longmapsto}
\def\goto{\mapsto}
\def\inf{\infty}
\def\imp{\Longrightarrow}
\def\iso{\cong}
\def\sub{\subseteq}
\def\sup{\supset}
\def\nul{\emptyset}
\def\smin{\setminus}
\def\Lam{\Lambda}
\def\Om{\Omega}
\def\om{\omega}
\def\lam{\lambda}
\def\al{\alpha}
\def\bt{\beta}
\def\del{\delta}
\def\Del{\Delta}
\def\eps{\epsilon}
\def\gam{\gamma}
\def\Gam{\Gamma}
\def\Sig{\Sigma}
\def\sig{\sigma}
\def\th{\theta}
\def\gm{{\mathfrak m}}

\begin{document}

\title{On Semigroups Generated by Two Consecutive Integers and
  Improved Hermitian Codes}
\author{Maria Bras-Amor\'os, Michael E. O'Sullivan,~\IEEEmembership{Member,~IEEE,}
\thanks{M. Bras-Amor\'os is with Universitat Aut\`onoma de Barcelona
  (e-mail: mbras@deic.uab.cat)}
\thanks{M. E. O'Sullivan is with San Diego State University (e-mail: mosulliv@sciences.sdsu.edu)}
}

\markboth{IEEE Transactions on Information Theory, VOL. X,
  NO. X}{BRAS-AMOR\'OS AND O'SULLIVAN: ON SEMIGROUPS GENERATED BY TWO
  CONSECUTIVE INTEGERS AND IMPROVED HERMITIAN CODES}

\maketitle

\begin{abstract}
Analysis of the Berlekamp-Massey-Sakata algorithm for decoding one-point
codes leads to two methods for improving code rate.
One method, due to Feng and Rao, removes parity checks that may be
recovered by their majority voting algorithm.  
The second method is to design the code to correct only those error
vectors of a given weight that are also geometrically generic.
In this work,  formulae are given for the redundancies 
of Hermitian codes optimized with respect to these
criteria
as well as the formula for the order bound on the minimum distance.
The results proceed from an analysis of numerical semigroups generated
by two consecutive  integers.
The formula for the redundancy of optimal Hermitian codes
correcting a given number of errors
answers  an open question stated by
Pellikaan and Torres in 1999.
\end{abstract}


\section*{Introduction}

\PARstart{N}{umerical} semigroups have proven to
be very useful in the study of
one-point algebraic-geometry codes.
On one hand the arithmetic of the numerical semigroup associated to
the one-point yields  a good bound---called the order bound---on
minimum distance \cite{FeRa:dFR,HoLiPe:agc,KiPe:telescopic}.
On the other hand, a close analysis of the numerical semigroup and
the decoding algorithm commonly used for one-point codes
shows that significant improvements in rate may be achieved while
maintaining a given error correction capability \cite{FeRa:improved}.
In this article we discuss the order bound and improvements to
the rate for codes constructed from Hermitian curves.

Let us briefly recall the definition of one-point algebraic geometry codes
and state the notation we will use.
Suppose $\F$ is a finite field, $F/\F$ a function field and $P$ a rational point of $F/\F$.
For $m\in\N_0$ let ${\mathcal L}(mP)$ be the ring of functions in $F$ having poles only at $P$ and of order at most $m$.
Let $v_P$ be the valuation of $F$ associated with $P$ and let
$\Lambda=\{-v_P(f): f\in \bigcup_m{\mathcal L}(mP)\}$.
$\Lambda$ is a {\it numerical semigroup}.
That is, a subset of $\N_0$, closed under summation, containing $0$
and with finite complement in $\N_0$.
It is called the {\it Weierstrass semigroup} associated to $P$.
Let $P_1, \dots, P_n$ be pairwise distinct rational points of $F/\F$
which are different from $P$  and let $\varphi$ be the map
$\bigcup_m{\mathcal L}(mP)\rightarrow\F^n$ such that $f\mapsto(f(P_1),\dots,f(P_n))$.
Suppose that $\Lambda=\{\lambda_0=0<\lambda_1<\lambda_2<\dots\}$.
The {\it $i$-th one-point algebraic-geometry code} associated with
$P$ and $P_1,\dots,P_n$ is $[\varphi({\mathcal L}(\lambda_i P))]^\perp$.
Naturally, the semigroup which will give us information about the one-point codes on $P$
will be the Weierstrass semigroup associated to $P$.

The {\it Hermitian curve} over $\F_{q^2}$, where
$q$ is a prime power,  is defined by its affine
equation
$x^{q+1}=y^{q}+y.$
It has a single point $P_\infty$
at infinity
and $q^3$ proper rational points $P_1,\dots,P_{q^3}$.
The ring of
functions on the curve with poles
only at $P_\infty$
is generated,
as a vector space over $\Fq$,
by
the set
$\{x^iy^j: j< q\}$.
Moreover,
$v_{P_\infty}(x)=-q$
and $v_{P_\infty}(y)=-q-1$.
Thus, the Weierstrass semigroup
at
$P_\infty$
is generated by $q$ and $q+1$.
{\it Hermitian codes}
are the one-point codes
defined on the Hermitian curve
associated with $P_\infty$ and $P_1,\dots,P_{q^3}$.
For details on the
Hermitian curve
and
the
Hermitian codes
we refer to \cite{Stichtenoth:hermite,HoLiPe:agc,Geil:Norm-trace-codes}.

The scope of this work is
to analyze some aspects
of Hermitian codes
based on the Weierstrass semigroup at $P_\infty$. 
Since the only thing we will be using about the Hermitian codes 
is that the associated numerical semigroup is generated by two
consecutive integers,
all the results can be stated more generally
for all those one-point codes for which the associated
semigroup is generated by two consecutive integers.
In Section~\ref{sec:enum} we analyze the enumeration of
 semigroups generated by two consecutive integers.
Then
we mention the known results on the sequence $\nu_i$
and the order
bound.
In Section~\ref{sec:red_st}
we give formulas for the
number of checks
of optimal 
codes correcting all errors of a given weight,
whenever the associated numerical semigroup is generated by two
consecutive integers.
In the case of Hermitian codes
this is the answer of an open question
stated in \cite{PeTo}.
In Section~\ref{sec:red_gen}
we give formulas for the
number of checks
of optimal codes correcting all {\it generic} errors of a given weight.

\section{On the enumeration and the $\nu$-sequence 
of semigroups generated by two consecutive integers}
\label{sec:enum}

We start this section with a small survey of the nomenclator and notations
we will use on numerical semigroups and, more specifically, those
numerical semigroups
generated by two consecutive integers.
Then we will analyze the enumeration
of the latter semigroups and give
the tools we will use in Section~\ref{sec:red_st} and
Section~\ref{sec:red_gen}.

\subsection{Semigroups Generated by Two Consecutive Integers}

By a {\it numerical semigroup}
we mean a subset of $\N_0$,
whose complement in $\N_0$ is finite and
which contains any sum of its elements.
Given a numerical semigroup $\Lambda$ we denote {\it gaps}
the elements in its complement in $\N_0$.
The {\it genus} $g$ of $\Lambda$ is the number of gaps
while its {\it conductor} $c$ is equal to the largest gap plus one.
The {\it enumeration} $\lambda$ of $\Lambda$ is the unique increasing bijective map
$\lambda:\N_0\longrightarrow\Lambda$. We say $\lambda_i$ to denote $\lambda(i)$.
Notice that if $\lambda_i$ is larger than or equal to the conductor or,
equivalently,  $i\geq c-g$, then
$\lambda_i=i+g$.

In this work we just deal with numerical semigroups
generated by two consecutive integers.
If the consecutive integers are $a,a+1$
then the numerical semigroup consists of any element
$ia+j(a+1)$ with $i,j\in\N_0$.
By properties
of semigroups generated by two integers \cite{HoLiPe:agc},
we know that the genus of this semigroup is $g=\frac{(a-1)a}{2}$ and its
conductor is $c=(a-1)a$.
Furthermore,
the semigroup generated by $a,a+1$
admits two alternative descriptions.
The first one is given by
the disjoint union
$0\sqcup\{a,a+1\}\sqcup\{2a,2a+1,2a+2\}\sqcup\dots\sqcup\{(a-2)a,(a-2)a+1,\dots,(a-2)a+a-2\}\sqcup\{i:i\geq
(a-1)a\}.$
The second one was proved in
\cite{GaRo:interval}
and it is given in the next lemma.
\begin{lemma}
\label{lemma:GaRo}
The numerical semigroup generated by $a,a+1$ is the set with all nonnegative integers
whose remainder when dividing by $a$
is at most the quotient.
\end{lemma}

\subsection{Enumeration}

As one can see
from Lemma~\ref{lemma:GaRo}, numerical semigroups generated by two consecutive integers
are highly related to
the set of pairs ${\mathcal P}=\{(x,y):x,y\in\N_0,y\leq x\}$.
In fact,
the numerical semigroup generated by $a,a+1$
is the image of the map
$$\begin{array}{crcl}
  \alpha_a:& {\mathcal P} & \rightarrow & \N_0\\
& (x,y) & \mapsto & ax+y\\
\end{array}
$$
It turns out that
this map is one-to-one
whenever $\alpha_a(x,y)$
is strictly less than
$a(a+1)$.
Indeed, if $l<a(a+1)$ and $(x,y)\in\alpha_a^{-1}(l)$ then
$x$ must be less than or equal to $a$ and $y$ must be strictly less than $a$.
So $x$ and $y$ are the quotient and the remainder of the
Euclidean division of $l$ by $a$, which are unique.
In particular,
$\alpha_a$ is one-to-one
whenever $\alpha_a(x,y)$
is less than or equal to the conductor
of the semigroup, which is $c=a(a-1)$.

Furthermore,
the total order
$$(x,y)<(x',y')\mbox{ if }\left\{\begin{array}{l}x<x',\\x=x'\mbox{ and }y<y',\end{array}\right.$$
is compatible with
the natural order of the semigroup
for all those values in the semigroup
which are less than $a(a+1)$.
That is,
for any $l,l'\in\Lambda$ with $l,l'<a(a+1)$,
then $l<l'\mbox{ if and only if }\alpha_a^{-1}(l)<\alpha_a^{-1}(l').$

Now, since $\sum_{j=0}^{k}j=\frac{k(k+1)}{2}$,
the sequence $a_k=\frac{k(k+1)}{2}$
is increasing and
$a_{k+1}-a_k=k+1$.
So any integer $i$ in $\N_0$ can be
written uniquely as
$i=\frac{x(x+1)}{2}+y$ for some $x\in\N_0$
and some $0\leq y\leq x$.
Thus, the map
$$\begin{array}{crcl}
  \beta:& {\mathcal P} & \rightarrow & \N_0\\
& (x,y) & \mapsto & \frac{x(x+1)}{2}+y\\
\end{array}
$$
is one-to-one everywhere
and it is also compatible with
the former total order.

As a conclusion,
and taking into consideration that
the genus and the conductor of
the numerical semigroup generated by $a,a+1$
are, respectively, $\frac{(a-1)a}{2}$
and $(a-1)a$,
one can see that the map
$\lambda:\N_0\longrightarrow\Lambda$ with
$$\lambda(i)=\left\{\begin{array}{ll}
\alpha o \beta^{-1}(i) & \mbox{ if }i\leq\frac{(a-1)a}{2},\\
i+\frac{(a-1)a}{2} & otherwise,\\
\end{array}\right.$$
is increasing and one-to-one.
Hence, it is exactly the enumeration
of the semigroup generated by $a,a+1$.

\subsection{The $\nu$-Sequence and the Order Bound}

Given a numerical semigroup $\Lambda$ with enumeration $\lambda$
define the sequence $\nu_i$ by
$$\nu_i=\lvert\{j\in\N_0:\lambda_i-\lambda_j\in\Lambda\}\rvert.$$
The sequence $\nu_i$ is used to define the {\it order bound} on the
minimum distance of one-point algebraic-geometry codes:
$$\delta_i=\min\{\nu_j:j>i\}.$$
The order bound, also known as Feng-Rao bound, is a lower bound
on the minimum distance of the $i$-th one-point code on $P$. In this
case the numerical semigroup
is the Weierstrass semigroup associated to $P$.
Details can be found in
\cite{FeRa:dFR,HoLiPe:agc,KiPe:telescopic}.

The Feng-Rao improved codes \cite{FeRa:improved}
are defined by means of the sequence $\nu_i$ as well.
First a set of functions
on the curve $\{z_i: i\in \N_0\}$
having only poles at $P$
is considered
such that
the valuation of $z_i$
at $P$ is $-\lambda_i$.
Now,
the Feng-Rao code designed to correct $t$ errors has
as parity checks the evaluation in certain points of the curve of functions $z_i$
for all $i$ with $\nu_i<2t+1$.

In this subsection we derive the sequence $\nu_i$ as well as the
order bound for numerical semigroups generated by two consecutive
integers. 
For Hermitian codes this information has appeared previously
(see \cite{PeTo,HoLiPe:agc,MuRa}). 
We choose to include our own proofs
since our methods are new and will be needed later in the analysis of
improved codes.

From now on, let $\Lambda$
be the semigroup generated by $a$ and $a+1$
and let $g$ and $c$ be respectively its genus and its conductor,
and let $\lambda$ be its enumeration.
In order to compute the values in the sequence $\nu_i$
we need to distinguish between those elements $\lambda_i\in\Lambda$
for which $\lambda_i=ax+y$ for unique nonnegative integers $x,y$ with $y\leq x$
from those
for which $x,y$ are not unique.

Let us denote by $\Lambda^x$ the subset of
$\Lambda$
containing
the elements $l=ax+y$
with $0\leq y\leq x$.
Then $l$ is uniquely expressible as $l=ax+y$
for nonnegative integers $x,y$ with $y\leq x$ if and only if
$l\in\Lambda^x\setminus(\cup_{x'\not=x}\Lambda^{x'})$.
Suppose $l=ax+y\in\Lambda^x$.
Then $l=a(x-1)+a+y$ and $l\in\Lambda^{x-1}$ if and only if
$a+y\leq x-1$, i.e., $y\leq x-a-1$.
Similarly,
$l=a(x+1)-a+y$ and $l\in\Lambda^{x+1}$ if and only if
$-a+y\geq 0$, i.e., $y\geq a$.
From this argument we have that
$ax+y$ with $y\leq x$ is in $\Lambda^x\setminus(\cup_{x'\not=x}\Lambda^{x'})$
if and only if $x-a\leq y\leq a-1$.


\begin{lemma}
\label{lemma: nu en Lambda isolada}
Let $\lambda_i\in\Lambda$ and suppose that the Euclidean division of
$\lambda_i$ by $a$ has quotient $x$ and remainder $y$.
If  $x-a\leq y\leq a-1$, then
$\nu_i=(x-y+1)(y+1)=xy-y^2+x+1.$
\end{lemma}

\begin{proof}
Suppose $\lambda_i=\lambda_j+\lambda_k$.
It is easy to check that if  $\lambda_i\in\Lambda^x\setminus(\cup_{z\not=x}\Lambda^z)$
for some $x$,
then $\lambda_j\in\Lambda^{x'}\setminus(\cup_{z\not=x'}\Lambda^z)$
and
$\lambda_k\in\Lambda^{x''}\setminus(\cup_{z\not=x''}\Lambda^z)$
for some $x',x''$.

So,
\begin{eqnarray*}
\nu_i&=&\lvert\{(x',y')\in{\mathcal P}: \lambda_i-ax'-y'\in\Lambda\}\rvert\\
&=&\lvert\{(x',y')\in{\mathcal P}:(x-x',y-y')\in{\mathcal P} \}\rvert\\
&=&\lvert\{(x',y')\in\N_0\times\N_0:\\
&&\phantom{mmm}x'\leq x,\ y'\leq y,\ y'\leq x',\ y'\geq x'-x+y\}\rvert\\
&=&\sum_{0\leq x'\leq x}\lvert\{y': \max\{0,y+x'-x\}\leq y'\leq \min\{y,x'\}\}\rvert.
\end{eqnarray*}

This last number is the number of integer points
inside a parallelogram with base $x-y+1$ and height $y+1$ (see
Figure~\ref{fig:paral}).
Hence it is equal to $(x-y+1)(y+1)$.

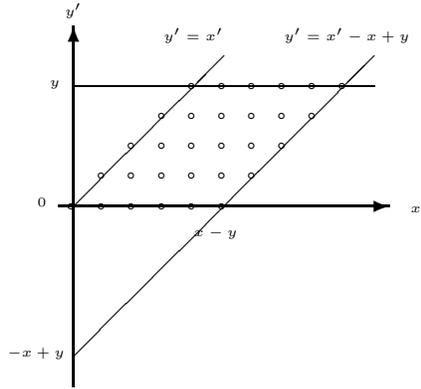
\begin{figure}[ht]
\setlength\unitlength{4mm}
\tiny
\begin{center}
\begin{picture}(16,15)
\thicklines
\put(3,1){\vector(0,1){12}}
\put(3,13.5){\makebox(0,0){$y'$}}
\put(2.5,7){\vector(1,0){11}}
\put(14.5,7){\makebox(0,0){$x'$}}

\thinlines
\put(3,7){\line(1,1){5}}
\put(6,12.5){$y'=x'$}
\put(3,2){\line(1,1){10}}
\put(10,12.5){$y'=x'-x+y$}
\put(3,11){\line(1,0){10}}
\put(0.8,2){$-x+y$}
\put(2.2,11){$y$}
\put(1.8,7){$0$}
\put(7,6){$x-y$}

\put(3,7){\makebox(0,0){\circle{0.2}}}
\put(4,7){\makebox(0,0){\circle{0.2}}}
\put(5,7){\makebox(0,0){\circle{0.2}}}
\put(6,7){\makebox(0,0){\circle{0.2}}}
\put(7,7){\makebox(0,0){\circle{0.2}}}
\put(8,7){\makebox(0,0){\circle{0.2}}}

\put(4,8){\makebox(0,0){\circle{0.2}}}
\put(5,8){\makebox(0,0){\circle{0.2}}}
\put(6,8){\makebox(0,0){\circle{0.2}}}
\put(7,8){\makebox(0,0){\circle{0.2}}}
\put(8,8){\makebox(0,0){\circle{0.2}}}
\put(9,8){\makebox(0,0){\circle{0.2}}}

\put(5,9){\makebox(0,0){\circle{0.2}}}
\put(6,9){\makebox(0,0){\circle{0.2}}}
\put(7,9){\makebox(0,0){\circle{0.2}}}
\put(8,9){\makebox(0,0){\circle{0.2}}}
\put(9,9){\makebox(0,0){\circle{0.2}}}
\put(10,9){\makebox(0,0){\circle{0.2}}}

\put(6,10){\makebox(0,0){\circle{0.2}}}
\put(7,10){\makebox(0,0){\circle{0.2}}}
\put(8,10){\makebox(0,0){\circle{0.2}}}
\put(9,10){\makebox(0,0){\circle{0.2}}}
\put(10,10){\makebox(0,0){\circle{0.2}}}
\put(11,10){\makebox(0,0){\circle{0.2}}}

\put(7,11){\makebox(0,0){\circle{0.2}}}
\put(8,11){\makebox(0,0){\circle{0.2}}}
\put(9,11){\makebox(0,0){\circle{0.2}}}
\put(10,11){\makebox(0,0){\circle{0.2}}}
\put(11,11){\makebox(0,0){\circle{0.2}}}
\put(12,11){\makebox(0,0){\circle{0.2}}}

\end{picture}
\end{center}
\caption{Parallelogram in proof of Lemma~\ref{lemma: nu en Lambda isolada}.}
\label{fig:paral}
\end{figure}

\end{proof}

To approach the case in which $\lambda_i=ax+y=ax'+y'$
with $x\neq x'$, $y\neq y'$,
we need a result from
\cite{Farran:symmetric}.
It says that if a numerical semigroup $\Lambda$ is such that its conductor $c$
is two times its genus,
then for all $\lambda_i\in\Lambda$
such that
$\lambda_i-c+1\in\Lambda$,
we have $\nu_i=\lambda_i-c+1$.
We already know that for the numerical semigroup generated
by $a,a+1$ the conductor is two times the genus.
Let us check that
if $\lambda_i\in\Lambda^{x}\cap\Lambda^{x+1}$
then
$\lambda_i-c+1\in\Lambda$.
Indeed, suppose  $\lambda_i\in\Lambda^{x}\cap\Lambda^{x+1}$.
Since
$\lambda_i\in\Lambda^{x+1}$,
$\lambda_i=(x+1)a+y$
with $y\leq x+1$.
Now, since
$\lambda_i\in\Lambda^x$
and $\lambda_i=xa+(a+y)$, we have
$a+y\leq x$.
Thus,
$\lambda_i-c+1=(x+1)a+y-a(a-1)+1=
a(x-a+2)+y+1$
with
$y+1\leq x-a+2$
and so
$\lambda_i-c+1\in\Lambda$.
Consequently,
if
$\lambda_i=ax+y=ax'+y'$
with $x\neq x'$, $y\neq y'$,
then
$\nu_i=\lambda_i-c+1$.

The next theorem is a consequence
of the former arguments.

\begin{theorem}
\label{th:nu}
Let $\lambda_i\in\Lambda$ and suppose that the Euclidean division of
$\lambda_i$ by $a$ has quotient $x$ and remainder $y$.
Then,
$$
\nu_i=\left\{\begin{array}{ll}
(x-y+1)(y+1)&\mbox{ if }-a+x\leq y \leq a-1,\\
\lambda_i-c+1&\mbox{otherwise.}\\
\end{array}\right.\\
$$
\end{theorem}

Once we have found a formula
for the values in the sequence $\nu_i$,
the next step is to find a formula
for the values of the order bound
defined as
$\delta_i=\min\{\nu_j:j>i\}.$
Notice that this definition has a lot to do with the increasingness
of the sequence $\nu_i$.

From Theorem~\ref{th:nu}
we deduce that
$\nu_i$ is quadratic in $y$ for the integers $i$
corresponding to the values $\lambda_i=ax+y$ inside $\Lambda^x$
with $-a+x\leq y \leq a-1$,
while it is increasing
elsewhere.
See Figure~\ref{fig:nuob4},
Figure~\ref{fig:nuob8},
Figure~\ref{fig:nuob16}.
By analyzing the parabola we see that
$\nu_i$ is increasing for
$y\leq\frac{x}{2}$
and decreasing for $y\geq\frac{x}{2}$, being symmetric
with respect to $y=\frac{x}{2}$.
In the case when $x<a$
all values $ax+y\in\Lambda^x$
satisfy
$-a+x\leq y \leq a-1$.
Then the first and last elements in $\Lambda^x$ (i.e. $y=0,y=x$)
have the same value for $\nu_i$, which is $x+1$ and which is minimal.
In the case when $x\geq a$,
the first element (i.e. $y=-a+x$) attains the minimal value for $\nu_i$, which is $ax-a^2+x+1$;
the second and last elements (i.e. $y=-a+x+1 ,y=a-1$) have the same value for $\nu_i$, which is $a(x-a+2)$ and
which is minimal if we take the first element away.
Thus,
\begin{itemize}
\item
If $x<a$ then
\begin{itemize}
\item
$\Lambda^x\cap\Lambda^{x'}=\emptyset$ for any $x'\neq x$
and
\begin{equation}\label{eq1}\min\{\nu_i:\lambda_i\in\Lambda^x\}=x+1,\end{equation}
\item
if $\lambda_i\in\Lambda^x$ and $\lambda_x\neq ax+x$ then
$$\min\{\nu_j:j>i \mbox{ and }\lambda_j\in\Lambda^x\}=x+1.$$
\end{itemize}
\item
If $a\leq x< 2a$ then
\begin{itemize}
\item $\Lambda^x\cap\Lambda^{x'}\neq\emptyset$,  $\Lambda^x\setminus(\cup_{x'\neq x}\Lambda^x)\neq \emptyset$, and
\begin{equation}\label{eq2}\begin{array}{l}\min\{\nu_i:\lambda_i=ax+y\in\Lambda^x,-a+x\leq y \leq a-1\}=\\\phantom{mmm}(a+1)x-a^2+1,\end{array}\end{equation}
\item
if $\lambda_i=ax+y\in\Lambda^x$ and $-a+x\leq y < a-1$ then
\begin{equation}\label{eq3}\begin{array}{l}\min\{\nu_j:j>i,\lambda_j=ax+y\in\Lambda^x,\\\phantom{mmm}-a+x\leq
  y \leq a-1\}=
a(x-a+2),\end{array}\end{equation}
\item
$\min\{\nu_i:\lambda_i\in\Lambda^x\cap\Lambda^{x+1}\}=\min\{\nu_i:a(x+1)\leq\lambda_i\leq ax+x\}=\nu_{\lambda^{-1}(a(x+1))}=a(x+1)-a(a-1)+1$,
\item
if $\lambda_i\in\Lambda^x\cap\Lambda^{x+1}$ and $\lambda_i\neq ax+x$, then\\
$\min\{\nu_j:j>i\mbox{ and }\lambda_j\in\Lambda^x\cap\Lambda^{x+1}\}=\lambda_{i+1}-c+1=\lambda_i-c+2.$
\end{itemize}
\item
If $x\geq 2a$ then
$\Lambda^x\setminus(\cup_{x'\neq x}\Lambda^x)= \emptyset$.
\end{itemize}
Finally, one can easily check the inequalities
\begin{itemize}
\item$\min\{\nu_i: \lambda_i\in\Lambda^{x-1}\cap\Lambda^{x}\}\leq\min\{\nu_i: \lambda_i\in{\Lambda}^x\setminus(\cup_{x'\neq x}\Lambda^{x'})\}\leq\min\{\nu_i: \lambda_i\in\Lambda^x\cap\Lambda^{x+1}\},$
\item$a(x-a+2)\leq\min\{\nu_i:\lambda_i\in\Lambda^x\cap\Lambda^{x+1}\},$
\item$\lambda_i-c+2\leq\min\{\nu_i: \lambda_i\in{\Lambda}^{x+1}\setminus(\cup_{x'\neq x+1}\Lambda^{x'})\},
\mbox{ for any }\lambda_i\in\Lambda^x\cap\Lambda^{x+1},\ \lambda_i\neq
ax+x.$
\end{itemize}

With these inequalities it is easy to prove the following theorem. We leave the details for the reader.

\begin{theorem}
\label{th:ob}
Let $\lambda_i\in\Lambda$ and suppose that the Euclidean division of
$\lambda_i$ by $a$ has quotient $x$ and remainder $y$.
Then,
$$
\delta_i=\left\{\begin{array}{ll}
x+1&\mbox{ if } x< a\mbox{ and }y\neq x,\\
x+2&\mbox{ if } x< a\mbox{ and }y= x,\\
a(x-a+2)&\mbox{ if } x\geq a \mbox{ and }-a+x\leq y <a-1,\\
\lambda_i-c+2&\mbox{ otherwise.}\\
\end{array}\right.\\
$$
\end{theorem}

The graphics in Figure~\ref{fig:nuob4},
Figure~\ref{fig:nuob8},
and Figure~\ref{fig:nuob16}
show the first values of $\nu_i$ and $\delta_i$ for the Hermitian codes over
$\F_{4^2}$, $\F_{8^2}$, and $\F_{16^2}$, 
respectively.

In fact, it is proven \cite{YaKu,HoLiPe:agc} that
for Hermitian
codes the order bound on the minimum distance is exactly the real
minimum distance of the codes.

\section{Minimizing redundancy}
\label{sec:red_st}

The decoding algorithm commonly used for one-point codes is an
adaptation of the Berlekamp-Massey-Sakata algorithm
\cite{Sakata} together with the majority voting algorithm of
Feng-Rao-Duursma \cite{FeRa,Duursma:maj,HoLiPe:agc}. By analyzing majority voting, one realizes
that only some of the parity checks are really necessary to perform
correction of a given number of errors. New codes can be defined
with just these few checks, yielding larger dimensions while keeping
the same correction capability as standard codes
\cite{FeRa:improved,HoLiPe:agc}. These codes are often called
Feng-Rao improved codes. The redundancy of standard one-point codes
correcting a given number $t$ of errors is
$$r(t)=\lambda^{-1}(\max\{i\in\N_0:\nu_i < 2t+1\})+1,$$
where the enumeration $\lambda$ and the sequence $\nu$
are derived from  the Weierstrass semigroup of the distinguished point.
The redundancy of the Feng-Rao improved codes
correcting the same number of errors is
$$\tilde{r}(t)=\lvert \{i\in\N_0:\nu_i < 2t+1\})\rvert.$$
This section is devoted to finding explicit formulae for
these redundancies
in the case when the associated Weierstrass semigroup is generated by
two consecutive integers $a,a+1$.
Recall that this is the case of Hermitian codes.

\begin{theorem}
Let $a>1$. Then,
{\tiny
\begin{eqnarray*}
r(t)
&=&
\left\{
\begin{array}{ll}
t(2t+1)&\mbox{ if }t\leq a/2,\\
(a^2-a)/2+(a+1)\lfloor\frac{2t}{a+1}\rfloor&\mbox{ if }a/2< t<a(\lfloor\frac{2t}{a+1}\rfloor+1)/2,\\
(a^2-a)/2+2t&\mbox{ if }t\geq a(\lfloor\frac{2t}{a+1}\rfloor+1)/2.\\
\end{array}\right.\\
\tilde{r}(t)
&=&
\left\{
\begin{array}{l}
t(2t+1)-\sum_{x'=\lceil2\sqrt{2t+1}-2\rceil}^{2t-1}(\lfloor
\sqrt{{x'}^2+4x'-8t}\rfloor+\delta_{x't})\\\hfill{\mbox{ if }t\leq a/2,}\\
(a^2-a)/2+(a+1)\lfloor\frac{2t}{a+1}\rfloor\\\phantom{mm}-\sum_{x'=\lceil2\sqrt{2t+1}-2\rceil}^{a-2+\lfloor\frac{2t}{a+1}\rfloor}
(\lfloor \sqrt{{x'}^2+4x'-8t}\rfloor+\delta_{x't})\\\hfill{\mbox{ if }a/2< t<a(\lfloor\frac{2t}{a+1}\rfloor+1)/2,}\\
(a^2-a)/2+2t-\sum_{x'=\lceil2\sqrt{2t+1}-2\rceil}^{a-1+\lfloor\frac{2t}{a+1}\rfloor}
(\lfloor\sqrt{{x'}^2+4x'-8t}\rfloor+\delta_{x't})\\\hfill{\mbox{ if }a(\lfloor\frac{2t}{a+1}\rfloor+1)/2\leq t\leq \frac{a(a+1)}{2},}\\
(a^2-a)/2+2t\\\hfill{\mbox{ if }t> \frac{a(a+1)}{2}},\\
\end{array}\right.\\
\end{eqnarray*}
}
where
{\tiny
$$\delta_{xt}=\left\{\begin{array}{ll}
1&\mbox{ if }x=\lfloor \sqrt{{x'}^2+4x'-8t}\rfloor \mbox{ mod } 2\\
0&\mbox{ if }x\neq\lfloor \sqrt{{x'}^2+4x'-8t}\rfloor \mbox{ mod } 2\\
\end{array}\right.=x+\lfloor \sqrt{{x'}^2+4x'-8t}\rfloor+1 \mbox{ mod }2.
$$}

\end{theorem}

\begin{proof}
By the arguments in the previous section,
the maximum non-gap whose $\nu$ is bounded by a certain constant
must be
1)
the last element in a parabola,
that is, $ax+x$ for some $x<a$ or $ax+a-1$ for some $x\geq a$;
2)
the first element in a parabola for some $x\geq a$, that is,
$ax+x-a$;
3)
some value in $\Lambda^{x'}\cap\Lambda^{x'+1}$
for some $x'$.
In case 1) and 2), $x$ is the largest integer such that
$\Lambda^x\setminus(\cup_{x'\neq x}\Lambda^{x'})\neq \emptyset$
and such that the minimum $\nu$ value in
$\Lambda^x\setminus(\cup_{x'\neq x}\Lambda^{x'})$
is at most $2t$.
That is, the corresponding parabola is not empty and its minimum value is at most $2t$.
In case 3),
if the largest integer $x$ such that
$\Lambda^x\setminus(\cup_{x'\neq x}\Lambda^{x'})\neq \emptyset$
and such that the minimum $\nu$ value in
$\Lambda^x\setminus(\cup_{x'\neq x}\Lambda^{x'})$
is at most $2t$, satisfies $x<2a-1$,
then $x'=x$. Otherwise,
$x'\geq x$.

By formulas \ref{eq1} and \ref{eq2}, the set of all minimum $\nu$ values among all non-empty parabolas
is
$$\begin{array}{lll}
M&=&\{\min\{\nu_i:\lambda_i\in\Lambda^{x'}\setminus(\cup_{x''\neq
  x'}\Lambda^{x''})\}:
\\&&\hfill{\Lambda^{x'}\setminus(\cup_{x''\neq x'}\Lambda^{x''})\neq\emptyset\}}\\
&=&\{x'+1:0\leq x'\leq a-1\}\cup\{(a+1)x'-a^2+1:
\\&&\hfill{a\leq x'<2a\}}\\
&=&\{z:1\leq z\leq a\}\cup\{z(a+1):1\leq z\leq a\}.
\end{array}$$
Now,
the maximum among these values which is at most $2t$
is

$
\begin{array}{l}
\max\{m\in M:m\leq 2t\}=\\\hfill{\left\{\begin{array}{ll}
2t &\mbox{ if } 2t\leq a,\\
\lfloor\frac{2t}{a+1}\rfloor(a+1) & \mbox{ if }a+1\leq 2t\leq a(a+1),\\
a(a+1) & \mbox{ if } 2t>a(a+1).\\
\end{array}\right.}
\end{array}
$

Therefore,

$$x=\left\{\begin{array}{ll}
2t-1&\mbox{ if }2t\leq a,\\
\lfloor\frac{2t}{a+1}\rfloor+a-1&\mbox{ if }a+1\leq 2t\leq a(a+1),\\
2a-1 & \mbox{ if } 2t>a(a+1).\\
\end{array}\right.$$

If $2t\leq a$ then $\Lambda^x\cap\Lambda^{x+1}=\emptyset$
and we are in case 1).
Otherwise, if $2t> a$ then
$\Lambda^x\cap\Lambda^{x+1}\neq\emptyset$.
If
$2t<a(x-a+2)$,
by formulas (\ref{eq2}) and (\ref{eq3}),
then we are in case 2).
Otherwise, we will be either in case 1) or 3).
Consequently,

{\small
\begin{eqnarray*}
r(t)
&=&
\left\{
\begin{array}{l}
\lambda^{-1}(ax+x)+1\\\hfill{\mbox{ if }2t\leq a,}\\
\lambda^{-1}(ax+x-a)+1\\\hfill{\mbox{ if }a< 2t<a(x-a+2),}\\
\lambda^{-1}(ax+a-1)+1+\mid\{\lambda_i\in\cup_{x'> x}\Lambda^{x'}:\nu_i\leq 2t\}\mid\\\hfill{\mbox{ if }2t\geq a(x-a+2).}\\
\end{array}\right.\\
\end{eqnarray*}
}

Replacing $x$ by its value and taking into consideration that
the value
$\nu_i$ increases constantly by one
within
$\{\lambda_i\in\cup_{x'> x}\Lambda^{x'}:\nu_i\leq 2t\}$,
we obtain

{\small
\begin{eqnarray*}
r(t)
&=&\left\{
\begin{array}{l}
t(2t+1)\\\hfill{\mbox{ if }t\leq a/2,}\\
(a^2-a)/2+(a+1)\lfloor\frac{2t}{a+1}\rfloor\\\hfill{\phantom{mmmmmmm}\mbox{ if }a/2< t<a(\lfloor\frac{2t}{a+1}\rfloor+1)/2,}\\
(a^2-a)/2+2t\\\hfill{\mbox{ if }t\geq a(\lfloor\frac{2t}{a+1}\rfloor+1)/2.}\\
\end{array}\right.\\
\end{eqnarray*}
}

For the result on $\tilde{r}(t)$
recall that the parabola $(x-y+1)(y+1)$
gives the values of $\nu_i$ for the non-gaps $\lambda_i=ax+y$ with $x-a\leq y\leq a-1$.
Fixed $x$, the maximum on $y$
of $(x-y+1)(y+1)$
is attained at $y=x/2$ and it is equal to
$x^2/4+x+1$.
From
the values $\lambda_i$
with $i< r(t)$
we want to take away all those
values whose corresponding $\nu_i$
is larger than $2t$.
Our first aim is to identify which parabolas
have nonempty intersection with the line at height $2t+1$.
That is, $x^2/4+x+1\geq 2t+1$.
Those are exactly the parabolas
for which $x\geq \lceil2\sqrt{2t+1}-2\rceil$.

Now, from each parabola we need to know which is the number of integers $y$
for which the $\nu_i$ corresponding to $\lambda_i=ax+y$
is at least $2t+1$.
Since the parabola
$(x-y+1)(y+1)$ is symmetric with respect to $y=x/2$,
there will be an odd number of such integers if $x$ is even
and an even number if $x$ is odd.
The real values $y$ where the parabola
equals $2t+1$
are given by the equation
$-y^2+xy+x+1=2t+1$,
and are exactly
$\frac{x\pm\sqrt{x^2+4x-8t}}{2}$.
Thus, the length
of the real interval
where the parabola is at least $2t+1$
is $\sqrt{x^2+4x-8t}$.
Now, from this interval we only want
its integer values.
It is easy to check that the number of such integers
is
$\lfloor\sqrt{{x}^2+4x-8t}\rfloor+\delta_{xt}$.

\end{proof}

\section{Minimizing redundancy for correcting generic errors}
\label{sec:red_gen}

In
\cite{O'Sullivan:hermite-beyond}
another improvement on one-point codes is described. Under the
Berlekamp-Massey-Sakata algorithm with majority voting, an error vector
whose weight is larger than half the minimum distance of the code is
often correctable.  In particular this occurs for {\it generic errors}
(also called independent errors in
\cite{Pellikaan:independent_errors,JeNiHo}), whose  technical
algebraic definition  can be found in \cite{BrOS:AAECC}.
Generic  errors of weight $t$ can be a very large proportion of all
possible errors of weight $t$, as in  the case of the examples worked out in
\cite{O'Sullivan:hermite-beyond}.
This suggests that a code be designed to correct only {\it
generic errors}  of weight $t$ rather than  all error words of
weight $t$.
Using this
restriction, one obtains new codes with much
larger dimension than that of standard one-point codes correcting
the same number of errors.
In  \cite{BrOS:AAECC}, the redundancy of standard
one-point codes correcting all generic errors of weight up to $t$
is shown to be
$$r^*(t)=\lambda^{-1}(\max( \Lambda \setminus \{ \lambda_i + \lambda_j:  i, j \geq t\})+1.$$
However, taking full advantage
of the Feng and Rao improvements due to the majority voting step \cite{FeRa:improved},
one can get optimal codes
correcting all generic errors of weight up to $t$
with redundancy
$$\tilde{r}^*(t)=\lvert \Lambda \setminus \{ \lambda_i + \lambda_j:  i, j \geq t\}\rvert.$$

This section is devoted to finding explicit formulae for
these redundancies.

It is easy to check that if $t$ is such that
$\lambda_t$ is larger than or equal to the conductor
then both $r^*(t)$ and $\tilde{r}^*(t)$ are equal to $\lambda_t+t$.
If $c$ is the conductor and $g$ is the genus, $\lambda_t\geq c$ is equivalent
to $t\geq c-g$.
More specifically, for the semigroup generated by $a,a+1$
this is equivalent to $t\in \Lambda^x$ for $x\geq a-1$.
In the next theorem we deal with the case when $t$ is strictly less than the conductor, that is,
when $t\in\Lambda^x$ with $x<a-1$.

\begin{theorem}
\label{theorem:red-hermite}
Suppose
$t=\frac{x(x+1)}{2}+y$
with $0\leq y\leq x<a-1$. That is,
$\lambda_t=xa+y$ with
$0\leq y\leq x<a-1$. Then,
\begin{eqnarray*}
r^*(t)&=&\left\{
\begin{array}{l}
2x^2+x\\\hfill{\mbox{ if } 2x<a,\ y=0,}\\
2x^2+3x+y+1\\\hfill{\mbox{ if } 2x<a,\ y>0,}\\
2xa+y-\frac{a^2-3a}{2}\\\hfill{\mbox{ if } 2x\geq a,\
y>2x-a+1,}\\
2xa+2y-\frac{a^2-a}{2}\\\hfill{\phantom{mmmmmm}\mbox{ if } 2x\geq a,\
y\leq 2x-a+1.}\\
\end{array}
\right.
\\
\tilde{r}^*(t)&=&\left\{
\begin{array}{l}
2x^2+x+3y\\\hfill{\mbox{ if } 2x<a,}\\
2xa+3y-2x-\frac{a^2-3a}{2}-1\\\hfill{\mbox{ if } 2x\geq a,\
y>2x-a+1,}\\
2xa+2y-\frac{a^2-a}{2}\\\hfill{\phantom{mmmmmm}\mbox{ if } 2x\geq a,\
y\leq2x-a+1.}\\
\end{array}
\right.
\end{eqnarray*}
\end{theorem}

\begin{proof}
We have
$\{\lambda_i+\lambda_j: i,j\geq t\}=
\{l\in\Lambda^{2x}: l\geq
2xa+2y\}
\cup
\{l\in\Lambda^{2x+1}: l\geq
(2x+1)a+y\}
\cup(\cup_{x'\geq 2x+2}\Lambda^{x'}).$
Notice that
$\{l\in\Lambda^{2x+1}: l<
(2x+1)a+y\}
\cap\Lambda^{2x+2}=\emptyset$
because $y<a$.
So,

$\Lambda\setminus\{\lambda_i+\lambda_j:i,j\geq t\}=
\{l\in\Lambda: l<2xa+2y\}
\sqcup
(\{l\in\Lambda^{2x+1}: l<(2x+1)a+y\}
\setminus\Lambda^{2x}).$

Let
\begin{eqnarray*}
A&=&\{l\in\Lambda: l<2xa+2y\},\\
B&=&\{l\in\Lambda^{2x+1}: l<(2x+1)a+y\}
\setminus\Lambda^{2x}.
\end{eqnarray*}

If $2x<a$ then
$\lvert A\rvert=\frac{2x(2x+1)}{2}+2y$ and
$\lvert B\rvert=y$ because
$\Lambda^{2x}\cap\Lambda^{2x+1}=\emptyset$.
So,

$
\begin{array}{lll}
\tilde{r}^*(t)&=&
\lvert\Lambda\setminus\{\lambda_i+\lambda_j:i,j\geq t\}\rvert\\
&=&\lvert A\rvert + \lvert B\rvert\\&=&2x^2+x+3y,\\
r^*(t)&=&\left\{\begin{array}{l}
\frac{2x(2x+1)}{2}=2x^2+x\\\hfill{ \mbox{\ if\ } y=0,}\\
\frac{(2x+1)(2x+2)}{2}+y=2x^2+3x+y+1\\\hfill{ \mbox{\ if\ } y>0.}\\
\end{array}\right.
\end{array}
$
If $2x\geq a$, then
all elements in
$\Lambda^{2x}$ are larger than the conductor
and
$\lvert A\rvert=2xa+2y-g=2xa+2y-\frac{a^2-a}{2}.$
In order to compute $\lvert B\rvert$, notice that
$\lvert \{l\in\Lambda^{2x+1}: l<(2x+1)a+y\}\rvert=y$,
while
$\lvert \Lambda^{2x}\cap\Lambda^{2x+1}\rvert
=2x-a+1.$
Now,
if $y>2x-a+1$, then
$\Lambda^{2x}\cap\Lambda^{2x+1}\subseteq
\{l\in\Lambda^{2x+1}: l<(2x+1)a+y\}$,
so $\lvert B\rvert=y-2x+a-1$ and
\begin{eqnarray*}
\tilde{r}^*(t)&=&\lvert A\rvert+\lvert B\rvert=2xa+3y-2x-\frac{a^2-3a}{2}-1,\\
r^*(t)&=&2xa+y-\frac{a^2-3a}{2}.
\end{eqnarray*}

Otherwise,
if $y\leq 2x-a+1$, then
$\Lambda^{2x}\cap\Lambda^{2x+1}\supseteq
\{l\in\Lambda^{2x+1}: l<(2x+1)a+y\}$,
so $\lvert B\rvert=0$ and
\begin{eqnarray*}
\tilde{r}^*(t)&=&\lvert A\rvert=2xa+2y-\frac{a^2-a}{2},\\
r^*(t)&=&\lvert A\rvert=2xa+2y-\frac{a^2-a}{2}.
\end{eqnarray*}
\end{proof}


\mut{

Let us see the behavior
of $r(t)$, $\tilde{r}(t)$,
$r^*(t)$ and  $\tilde{r}^*(t)$
for some examples of Hermitian curves.
Notice that in general
$r(t)>\tilde{r}(t)
>r^*(t)>\tilde{r}^*(t)$
and that the differences
are largest for small
values of $t$.

\begin{center}
\input{../GraphDirectory/red-hermite}
\end{center}

Note that
the graph of
$r(t)$, $\tilde{r}(t)$,
$r^*(t)$ and  $\tilde{r}^*(t)$
seems to become more regular as the cardinal of the
finite field increases.
}

\begin{figure}
\resizebox{\columnwidth}{!}{
\def\simbolnu{\circle{3}}
\def\simbolobound{\makebox(0,0){\resizebox{5\unitlength}{5\unitlength}{$\times$}}}

}
\caption{Graph of $\nu_i$ and $\delta_i$ for the Hermitian code over $\F_{4^2}$.}
\label{fig:nuob4}
\end{figure}

\begin{figure}
\resizebox{\columnwidth}{!}{
\def\simbolnu{\circle{3}}
\def\simbolobound{\makebox(0,0){\resizebox{5\unitlength}{5\unitlength}{$\times$}}}
%
}
\caption{Graph of $\nu_i$ and $\delta_i$
for the Hermitian code over $\F_{8^2}$.}
\label{fig:nuob8}
\end{figure}

\begin{figure}
\resizebox{\columnwidth}{!}{
\def\simbolnu{\circle{3}}
\def\simbolobound{\makebox(0,0){\resizebox{5\unitlength}{5\unitlength}{$\times$}}}
%
}
\caption{Graph of $\nu_i$ and $\delta_i$
for the Hermitian code over $\F_{16^2}$.}
\label{fig:nuob16}
\end{figure}

\begin{figure}
\resizebox{\columnwidth}{!}{
\def\simbolnu{\circle{3}}
\def\simbolobound{\makebox(0,0){\resizebox{5\unitlength}{5\unitlength}{$\times$}}}
%
}
\caption{Graph of $\nu_i$ and $\delta_i$
for the Hermitian code over $\F_{32^2}$.}
\label{fig:nuob32}
\end{figure}

\def\cprime{$'$}



\end{document}